\documentclass[sigconf]{acmart}

\AtBeginDocument{%
  }

\setcopyright{acmlicensed}
\copyrightyear{2024}
\acmYear{2024}
\acmDOI{XXXXXXX.XXXXXXX}

\acmConference[ISACom'24]{3rd ACM MobiCom Workshop on Integrated Sensing and Communication Systems for 6G}{November 18, 2024}{Washington, D.C., USA}

\acmISBN{978-1-4503-XXXX-X/18/06}

\usepackage{subfigure}
\usepackage{caption}
\usepackage{subcaption}
\copyrightyear{2024}
\acmYear{2024}
\setcopyright{acmlicensed}\acmConference[ACM MobiCom '24]{The 30th Annual International Conference on Mobile Computing and Networking}{November 18--22, 2024}{Washington D.C., DC, USA}
\acmBooktitle{The 30th Annual International Conference on Mobile Computing and Networking (ACM MobiCom '24), November 18--22, 2024, Washington D.C., DC, USA}
\acmDOI{10.1145/3636534.3698221}
\acmISBN{979-8-4007-0489-5/24/11}

\begin{document}

\title{Access Point  Deployment for Localizing Accuracy and User Rate in Cell-Free Systems}

\author{Fanfei Xu}
\affiliation{%
	\institution{Southeast University School of Information Science and Engineering}
	\city{Nanjing 210096}
	\country{China}}

\author{Shengheng Liu}
\authornote{Both authors are also affiliated to the Purple Mountain Laboratories, Nanjing 211111, China.}
\affiliation{%
	\institution{Southeast University School of Information Science and Engineering}
	\city{Nanjing 210096}
	\country{China}}
\email{s.liu@seu.edu.cn}

\author{Zihuan Mao}
\affiliation{%
	\institution{Southeast University School of Information Science and Engineering}
	\city{Nanjing 210096}
	\country{China}}
	
\author{Shangqing Shi}
\affiliation{%
	\institution{Southeast University School of Information Science and Engineering}
	\city{Nanjing 210096}
	\country{China}}

\author{Dazhuan Xu}
\affiliation{%
	\institution{Purple Mountain Laboratories}
	\city{Nanjing 211111}
	\country{China}}

\author{Dongming Wang}
\affiliation{%
	\institution{Southeast University School of Information Science and Engineering}
	\city{Nanjing 210096}
	\country{China}}

\author{Yongming Huang}
\authornotemark[1]
\affiliation{%
	\institution{Southeast University School of Information Science and Engineering}
	\city{Nanjing 210096}
	\country{China}}

\renewcommand{\shortauthors}{Xu et al.}

\begin{abstract}
Evolving next-generation mobile networks is designed to provide ubiquitous coverage and networked sensing. With utility of multi-view sensing and multi-node joint transmission, cell-free is a promising technique to realize this prospect. This paper aims to tackle the problem of access point (AP) deployment in cell-free systems to balance the sensing accuracy and user rate. By merging the D-optimality with Euclidean criterion, a novel integrated metric is proposed to be the objective function for both max-sum and max-min problems, which respectively guarantee the overall and lowest performance in multi-user communication and target tracking scenario. To solve the corresponding high dimensional non-convex multi-objective problem, the Soft actor-critic (SAC) is utilized to avoid risk of local optimal result. Numerical results demonstrate that proposed SAC-based APs deployment method achieves $20\%$ of overall performance and $120\%$ of lowest performance.

%	Soft actor-critic (SAC), a deep reinforcement learning (DRL) algorithm with additive maximizing entropy term is employed to solve the original non-convex multi-objective problem having high-dimensional variables. Numerical results demonstrate proposed SAC-based APs deployment method achieving higher user communication rate and target localizing accuracy than other traditional DRL algorithms.
	
	%and the fairness among users and moving target trajectory.

	%evaluate the ISAC performance of APs deployment. 

	%, which is then optimized as max-min and max-sum problem to guarantee  the fairness between user ra and target sensing. 

	%be the objective function for both max-sum and max-min problem, which guarantees systems overall performance and fairness service among users and moving target trajectory. 
%	Soft actor-critic (SAC), a deep reinforcement learning (DRL) algorithm with additive maximizing entropy term is employed to solve the original non-convex multi-objective problem having high-dimensional variables. Numerical results demonstrate proposed SAC-based APs deployment method achieving higher user communication rate and target localizing accuracy than other traditional DRL algorithms.
	
\end{abstract}


%\ccsdesc[500]{Do Not Use This Code~Generate the Correct Terms for Your Paper}
%\ccsdesc[300]{Do Not Use This Code~Generate the Correct Terms for Your Paper}
%\ccsdesc{Do Not Use This Code~Generate the Correct Terms for Your Paper}
%\ccsdesc[100]{Do Not Use This Code~Generate the Correct Terms for Your Paper}

\received{XX XXXX 2024}
\received[revised]{XX XXXX 2024}
\received[accepted]{XX XXXX 2024}

\maketitle

\section{Introduction}
The integrated sensing and communication (ISAC) is expected to become a key usages scenario of the future six-th generation (6G) networks, which is mentioned in newest recommendation concerned with framework and objective of future network of the  International Telecommunication Union \cite{ITU}. Telecommunication base stations, as a widely deployed infrastructure, can significantly enhance situational awareness when equipped with radar sensing capabilities, enabling a variety of novel applications, such as low-altitude drone intrusion detection and accurate virtual environment construction for digital twins \cite{digitaltwin}. Moreover, as wireless communication networks continue to evolve towards higher frequency bands, more spectrum resources can be obtained to satisfy the high bandwidth requirements of sensing functions. However, high-frequency signals with poor diffraction and high attenuation, makes traditional single station sensing performance significantly limited to shadow effects of obstructions. 

Cell-free networks \cite{ngo,xiaohu}, an innovative architecture for 6G networks, may tackle this issue. Unlike conventional cellular networks, where each user equipment (UE) is served by a single base station within a designated cell, cell-free networks deploy a large number of distributed access points (APs) that collaboratively serve all users within the area. Widely-spread APs connected to a central processing unit (CPU) enable seamless sensing and ubiquitous coverage by use of key techniques such as dynamic user association and APs deployment \cite{survey}. Different APs location seriously affects the channel condition, intuitively, to fully leverage the advantages such as multi-view sensing information and joint transmission of the cell-free ISAC systems, investigating optimal APs deployment is vitally necessary. 
%Intuitively, adjusting the AP deployment 
%Implementing ISAC function on existing cell-free networks can fully multi-view sensing information, avoid blind spots and thus dramatically enhance situational awareness.To fully leverage the advantages of the cell-free systems, investigating optimal APs deployment strategy is necessary.

Mathematical solutions such as vector quantization and gradient descent \cite{vector,hamid,fanfei} were proposed to optimize the two dimension (2D) location of APs to improve of spectral efficiency in cell-free networks. On-demand service capability concerned with sum throughput was achieved by solving APs deployment \cite{yinghui} based on multiple linear regression model. % Joint AP deployment and synchronization of distributed APs affecting phase accuracy are also important for cooperative communication \cite{depsynch}. In addition to systems performance, APs deployment topology \cite{topology} also significantly determines the capital and operational expenditure, which is the essential factor for operator to practically apply cell-free systems. 
Moreover, APs deployment not only affects communication performance, but also impacts sensing performance. Cram\'er-Rao lower bound (CRLB) was developed \cite{mimoradar} for target velocity estimation in multiple-input multiple output (MIMO) radar, shown that the antenna placement affects the estimation accuracy significantly. Also in MIMO radar systems \cite{localization}, CRLB for target localization in both coherent and non-coherent processing was developed. Additionally, based on the best unbiased linear unbiased estimator it derived, a closed-form localization estimation that revealed the relationship between antennas location, target location, and localization accuracy was provided. Furthermore, geometry gain of antennas deployment for target localizing is \cite{geometry} analyzed  in MIMO radar systems. In summary, substantial researches have separately analyzed the impacts of APs deployment on communication and sensing performance. However, the area of jointly considering both of them still remains blanket.

Thus, to simultaneously measure sensing and communication performance, in this work we propose a unified evaluation metric in cell-free ISAC systems, merging user rate with localizing accuracy \cite{surveyisac} derived from Euclidean distance and D-optimal criterion, and utilize it as the objective function for APs deployment optimization. In addition, considering fairness for all UEs and localizing accuracy throughout the target moving trajectory, we provide the deployment results of both max-sum and max-min problem. Due to the non-convex and high-dimensional property of the original multi-objective problem, mathematical algorithms are challenging to solve it. Soft actor-critic (SAC) \cite{SAC}, a deep reinforcement learning (DRL) based APs deployment method is proposed, with the utility of additive maximum AP deployment entropy term to avoid local optimum. Numerical results show the superior performance of our proposed SAC-based deployment method compared with other DRL algorithms such as deep deterministic policy gradient (DDPG) and twin-delayed DDPG (TD3).
\begin{figure}[h]
	\centering
	\includegraphics[scale=0.45]{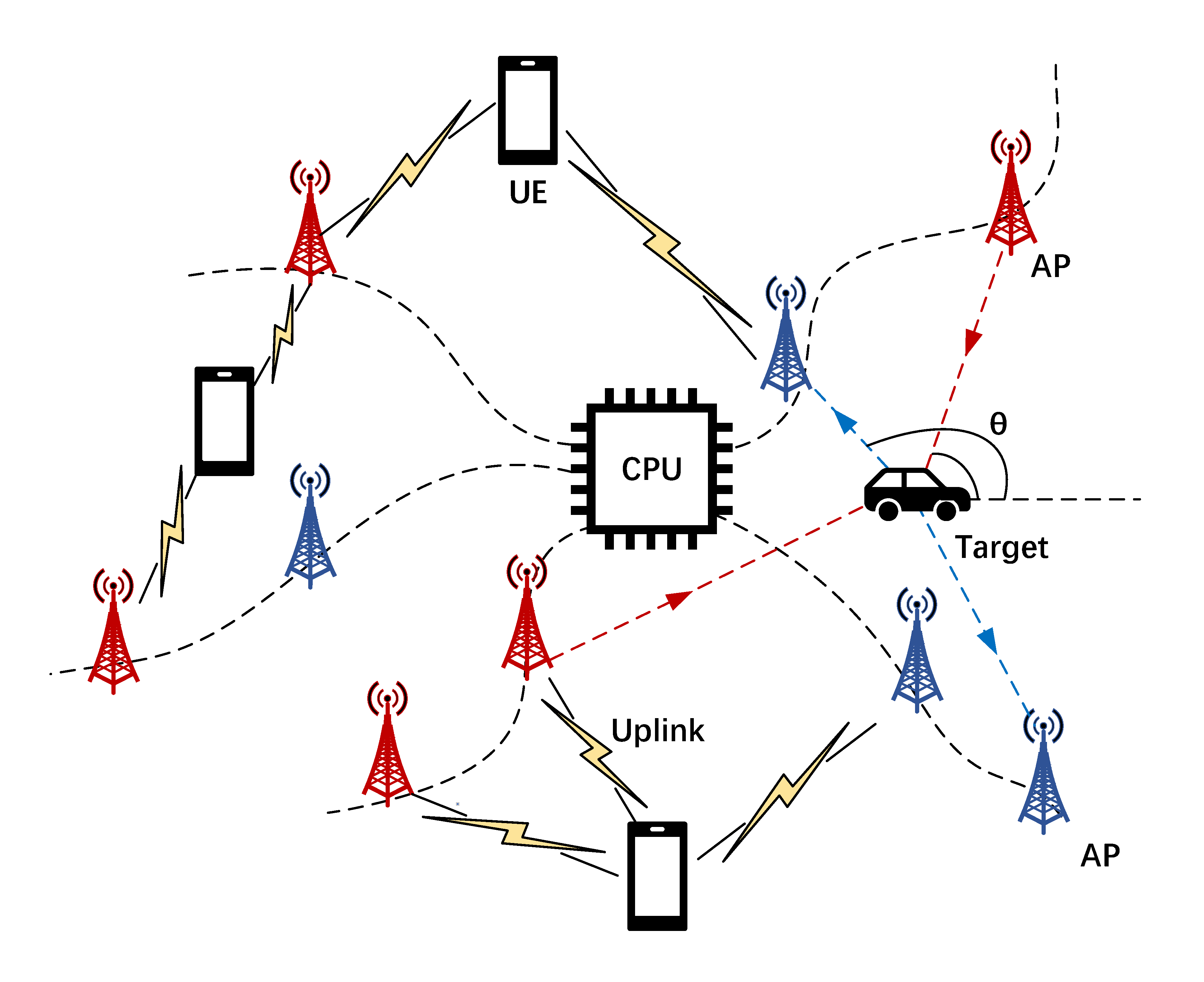}
	\caption{Cell-free ISAC systems.}
	\label{scenario}
\end{figure}
\vspace{-5mm}
\section{ISAC Signal Model and Problem Formulation}
As illustrated in Fig.~\ref{scenario}, we consider a cell-free ISAC system consists of $M$ single-antenna transmitter APs and $N$ single-antenna receiver APs which collaboratively serve $K$ single-antenna UEs and estimate position of one target moving with specific trajectory. The 2D position of them is denoted by $\mathbf{t}_m = [x^{\text{t}}_{m},y^{\text{t}}_{m}]$, $\mathbf{r}_n = [x^{\text{r}}_{n},y^{\text{r}}_{n}]$, $\mathbf{u}_k = [x_k, y_k]$, $\mathbf{p}=[x^{\text{p}}, y^{\text{p}}]$ respectively, where $m \in \mathbb{M}=\{1,\dots,M\}$, $n \in \mathbb{N}=\{1,\dots,N\}$, $k \in \mathbb{K}=\{1,\dots,K\}$.
\subsection{Fisher Information}
When system is executing sensing function module, $M$ transmitting APs send probe signal and $N$ receiving APs capture the echo from target. In this distributed detection system, the low-pass equivalent of the narrow-band signal transmitted from $i$-th APs at time $t$ is represented as $\sqrt{E/M}s_i(t)$, where $E$ denotes the total transmission energy.
We assume that the transmission signal is uncorrelated in any time delay,
\begin{equation}
	\int_{T}s_i(t)s_j^{*}(t-\tau) dt \approx 
	\begin{cases}
		1, & \text{if} \quad i=j \\ 
		0, &\text{if} \quad i \ne j,
 	\end{cases}
\end{equation}
where $(\cdot)^*$ represent the conjugate operator. In addition, the signal is normalized in the whole signal processing interval $T$, i.e., $\int_{T} |s_i(t)|^2 dt = 1$.

Non-coherent processing, a more practical operation is selected due to its low requirement of time synchronization compared with coherent processing. The echo signal accepted at $n$-th receiving AP is denoted by
\begin{equation}
	y_n(t) = \sum_{m=1}^{M}\eta_{mn}s_m(t-\tau_{mn}) + w_n(t),
\end{equation}
where $w_n(t) \stackrel{\text{i.i.d.}}{\sim} \mathcal{CN}(0,1)$ represents additive Gaussian noise, $\mathbf{\eta} = [\eta_{11}, \eta_{12},\dots,\eta_{mn},\dots,\eta_{MN}]^\text{T}$ is the coefficient referring to target reflection and channel propagation fading. $\tau_{mn}$ denotes time delay of signal transmission and reflection between $m$-th transmitting AP and $n$-th receiving AP
\begin{equation}
	\tau_{mn} = \frac{d_m+d_n}{c},
\end{equation}
where $c$ is speed of light and $d_m = \|\mathbf{t}_m- \mathbf{p}\|_2$, $d_n = \|\mathbf{r}_n-\mathbf{p}\|_2$ denote the distance between target $\mathbf{p}$ and transmitter or receiver AP.

%当无蜂窝网络执行感知功能时，$M$个坐标为$\mathbf{t}_m$的AP呈发射状态，$N$个坐标为$\mathbf{r}_n$的AP呈接收状态，并由以上的收发架构组成分布式的组网探测网络。
%We consider the two-dimensional ellipse  time of arrival (ToA) estimation formed by the $m$-th transmitting AP and the $n$-th receiving AP
%\begin{align}
%	\hat{\tau}_{mn}(\mathbf{p}) = \tau_{mn}+e_{mn},
%\end{align}
%where $e_{mn}$ is the error of ToA and , respectively. Then the estimation of distance can be expressed as
%\begin{align}
%	\hat{d}_{mn}(\mathbf{p}) = d_m + d_n +{\rm{c}}e_{mn}.
%\end{align}
%The vector form of distance estimation is
%\begin{align}
%	\tilde{\mathbf{d}} = \mathbf{d}(\mathbf{p})+\mathbf{w} = \left[d_1,d_2,\cdots,d_{MN}\right]^{\text{T}}+\left[w_1,w_2,\cdots,w_{MN}\right]^{\mathsf{T}}.
%\end{align}
%%其中$\mathbf{n}$可视为距离估计时的噪声，此时有距离估计值的协方差矩阵为
%The covariance matrix of the distance estimation is
%\begin{align}
%	\boldsymbol{\Sigma} = {\rm{E}} \{\mathbf{w}\mathbf{w}^{\text{T}}\} = {\rm{diag}}\left\{\sigma_1^2,\cdots,\sigma_{MN}^2\right\}.
%\end{align}
%当ToA估计值的误差为高斯噪声时，目标探测的费雪信息矩阵为

As for the localizing accuracy, we utilize D-optimal criterion to evaluate the performance of parameter estimation. The product of the CRLB matrix eigenvalues (or determinant) is minimize, which is equivalent to minimize the area of the Elliptical error probable. Fisher information matrix (FIM) is the inverse of CRLB matrix, therefore, minimizing the determinant of CRLB matrix converts to maximizing the FIM. 
The FIM of the estimation for the parameter vector $\mathbf{\vartheta}$ by use of the measurement vector $\mathbf{x}$ can be expressed as:
\begin{equation}
	\mathbf{\Phi} = \rm{E}\left\{\left[\frac{\partial}{\partial\mathbf{\vartheta}}\ln f(\mathbf{x}|\mathbf{\vartheta}) \right]\left[\frac{\partial}{\partial\mathbf{\vartheta}}\ln f(\mathbf{x}|\mathbf{\vartheta}) \right]^{\text{T}}\right\}.
\end{equation}

To be specific, when the error of the estimation is Gaussian noise, the FIM for in cell-free ISAC system, non-coherent sensing FIM for target detection is equal to
\begin{align}
	\boldsymbol{\Phi} = 
	\begin{bmatrix}
		\phi_{11} & \phi_{12}    \\
		\phi_{21} & \phi_{22}
	\end{bmatrix}
	=\mathbf{J}_0^{\rm{T}}\boldsymbol{\Sigma}^{-1}\mathbf{J}_0,
\end{align}
where $\mathbf{J}_{0}$ is the Jacobian matrix of the target localizing at $\mathbf{p}=[x^{\text{p}}, y^{\text{p}}]$
\begin{align}
	\mathbf{J}_0 = \left[
	\left(\mathbf{\alpha}_1^t+\mathbf{\alpha}_1^r\right)  \left(\mathbf{\alpha}_1^t+\mathbf{\alpha}_2^r\right) \cdots \left(\mathbf{\alpha}_1^t+\mathbf{\alpha}_N^r\right) \cdots \left(\mathbf{\alpha}_M^t+\mathbf{\alpha}_N^r\right)
	\right]^{\text{T}},
\end{align}
where $\mathbf{\alpha}_{m}^{t}=\left[\cos\theta_m^t \sin\theta_m^t\right]^{\text{T}}$,$\mathbf{\alpha}_{n}^{r}=\left[\cos\theta_n^r \sin\theta_n^r\right]^{\text{T}}$, $\theta$ represents the bearing angle of the AP relative to the target, measured from the horizontal axis. According to \cite{geometry}, the expression of determinant of the angular FIM is
\begin{equation}
	\begin{aligned}
				\vert\boldsymbol{\Phi}\vert &= \left\{\sum_{m=1}^{M}\sum_{n=1}^{N}\left(\cos{\theta_m^t}+\cos{\theta_n^r}\right)^2\sum_{m=1}^{M}\sum_{n=1}^{N}\left(\sin{\theta_m^t}+\sin{\theta_n^r}\right)^2\right.\\
				&\left.-\left[\sum_{m=1}^{M}\sum_{n=1}^{N}\left(\cos{\theta_m^t}+\cos{\theta_n^r}\right)\left(\sin{\theta_m^t}+\sin{\theta_n^r}\right)\right]^2\right\}.
		\end{aligned}
\end{equation}

Next, we transform the determinant of the angular FIM into two-dimensional Cartesian coordinates, and determine optimal APs deployment positions based on the maximum value of the FIM determinant.
\begin{equation}
	\begin{aligned}
		\vert\boldsymbol{\Phi}\vert \!\!&=\!\! \left\{\sum_{m=1}^{M}\sum_{n=1}^{N}\left(\frac{x^p\!-\!x^t_{m}}{\|\mathbf{p}\!-\!\mathbf{t}_m\|_2}\!\!+\!\!\frac{x^p\!-\!x^r_{n}}{\|\mathbf{p}\!-\!\mathbf{r}_n\|_2}\right)^2\!\!\sum_{m=1}^{M}\sum_{n=1}^{N}\left(\frac{y^p \!-\! y^t_{m}}{\|\mathbf{p}\!-\!\mathbf{t}_m\|_2}\!\!+\!\!\frac{y^p\!-\!y^r_{n}}{\|\mathbf{p}\!-\!\mathbf{r}_n\|_2}\right)^2\right.\\
		&\left.-\left[\sum_{m=1}^{M}\sum_{n=1}^{N}\left(\frac{x^p\!-\!x^t_{m}}{\|\mathbf{p}\!-\!\mathbf{t}_m\|_2}\!\!+\!\!\frac{x^p\!-\!x^r_{n}}{\|\mathbf{p}\!-\!\mathbf{r}_n\|_2}\right)\left(\frac{y^p\!-\!y^t_{m}}{\|\mathbf{p}\!-\!\mathbf{t}_m\|_2}\!\!+\!\!\frac{y^p\!-\!y^r_{n}}{\|\mathbf{p}\!-\!\mathbf{r}_n\|_2}\right)\right]^2\right\}.
	\end{aligned}
\end{equation}

\subsection{User Sum Rate}

An cell-free uplink communication model is studied in this section, where all $M+N$ APs simultaneously accept signals from all $K$ UEs.The uplink receiving signal at $l$-th AP  is 

\begin{equation}
	y_l = \sum_{k=1}^{K}\sqrt{\rho}h_{lk}x_k + w_l,
\end{equation}
where$\sqrt{\rho}$, $x_k$ is transmitter power and data symbol of $k$-th UE, $l \in \{1,2,\dots,M+N\}$. The received signal of all $M+N$ APs can be represented as
\begin{equation}
	\mathbf{y} = \sqrt{\rho}\mathbf{H}\mathbf{x} + \mathbf{w},
\end{equation}
where $\mathbf{x} = [x_1,x_2,\dots,x_K]^{\text{T}}$, $\mathbf{w}=[w_1,w_2,\dots,w_{M+N}]^{\text{T}}$and $\mathbf{H}[l,k] = h_{lk}, \mathbf{H} \in \mathbb{C}^{(M+N)\times K}$ is channel coefficients matrix. 
The narrow-band fading channel coefficient is  written as	$h_{lk}=\sqrt{\beta_{lk}}g_{lk}$, where $g_{lk} \stackrel{\text{i.i.d.}}{\sim}\mathcal{C}\mathcal{N}(0,1)$  small fading coefficients, and $\beta_{lk}$ is large fading coefficients 
\begin{equation}
	\beta_{lk} = 
	\begin{cases}
		\frac{d_0}{\|\mathbf{t}_l - \mathbf{u}_k\|_2^{2}}, & if \quad l \le M\\
		\frac{d_0}{\|\mathbf{r}_l - \mathbf{u}_k\|_2^{2}}, & if \quad M < l \le M+N,
	\end{cases}
\end{equation}
where $d_0$ is a constant denoting the reference distance.

In cell-free systems, signals accepted by distributed APs	at various  geographic position can be collectively  processed in CPU. For instance, zero forcing reception is employed to mitigate  interference from multiple UEs. Then, the processed received signal is expressed as
\begin{equation}
	\hat{\mathbf{y}} = \left(\mathbf{H}^{\text{H}}\mathbf{H}\right)^{-1}\mathbf{H}^{\text{H}}\mathbf{y}.
\end{equation}

The asymptotic SNR can be expressed as
\begin{equation}
	\frac{1}{M+N}\mathbf{SNR}_k\xrightarrow[(M+N)\rightarrow\infty]{a.s.}\rho\overline{\beta}_{k},
\end{equation}
where $\overline{\beta}_{k} \triangleq \lim\limits_{(M+N)\rightarrow\infty}\frac{1}{M+N}\sum_{l}\beta_{lk}$. For simplification of derivation, we consider the transmit power to be unity for all UEs. We assume no shadow fading in the large-scale fading coefficient $\beta_{lk}$,  and constant $d_0$ is 1. These assumptions do not affect the conclusions of subsequent methods. At this point, the  SNR for $k$ UE can be expressed by use of Euclidean distance criterion:
\begin{equation}
	\mathbf{SNR}_k = \left(\sum_{m=1}^{M}\frac{1}{\|\mathbf{u}_k - \mathbf{t}_m\|_2^{2}}+\sum_{n=1}^{N}\frac{1}{\|\mathbf{u}_k - \mathbf{r}_n\|_2^{2}}\right).
\end{equation}

The sum communication rate of cell-free systems is
\begin{equation}
	\begin{aligned}			
	\sum_{k=1}^{K} R_k & = \sum_{k=1}^{K} \log\left(1+\mathbf{SNR}_k\right) \\ & = \sum_{k=1}^{K}\log\left(\sum_{m=1}^{M}\frac{1}{\|\mathbf{u}_k - \mathbf{t}_m\|_2^{2}}+\sum_{n=1}^{N}\frac{1}{\|\mathbf{u}_k - \mathbf{r}_n\|_2^{2}}\right).
	\end{aligned}
\end{equation}
\subsection{Deployment Problem Formulation}

In the context of cell-free ISAC systems, the comprehensive consideration of communication and sensing performance can be modeled as a multi-objective optimization problem
\begin{equation}\label{optimization}
	\begin{array}{ll}
		\max\limits_{\mathbf{t}, \mathbf{r}} & U(\mathbf{t}, \mathbf{r}) = \left( U_{1}, U_2 \right) \\
		\text { s.t. } &x_{\text{min}} \leq x_i(x) \leq x_{\text{min}}, \quad i=1, \ldots, M+N \\
		& y_{\text{min}} \leq y_i(x) \leq y_{\text{min}}, \quad i=1, \ldots, M+N,
	\end{array}
\end{equation}
constraints in (\ref{optimization}) mean that all APs must be deployed within a specific area.

Directly solving the Pareto front of multi-objective problems is highly challenging. Therefore, we transform the original problem into a single-objective optimization problem for solution.

Due to the dimensional disparity between localizing accuracy and communication rate, traditional weighted sum approaches struggle to achieve balanced optimization between the two. Therefore, we adopt a multiplication method to transform the original problem for resolution, thereby achieving a better trade-off between communication and sensing performance.

We firstly consider the maximizing sum of communication capacity and sensing accuracy during the target moving trajectory. In addition, to ensure the uniform service for all UEs and sensing accuracy throughout the entire trajectory period, we also consider the maximization of minimum capacity and accuracy. Then, the objective function can be represented as
%\begin{equation}
%	\begin{array}{ll}
%		\max\limits_{\mathbf{t}, \mathbf{r}} & \left(\sum_{k=1}^{K}R_k,\sum \limits_{\mathbf{p}(t)} \vert\boldsymbol{\Phi}\vert \right) \\
%		\text { s.t. } &x_{\text{min}} \leq x_i(x) \leq x_{\text{min}}, \quad i=1, \ldots, M+N \\
%		& y_{\text{min}} \leq y_i(x) \leq y_{\text{min}}, \quad i=1, \ldots, M+N,
%	\end{array}
%\end{equation}
\begin{equation}
	U\left( \mathbf{t}, \mathbf{r} \right) = 
	\begin{cases}
		\sum_{k=1}^{K}R_k/Q\cdot\sum \limits_{\mathbf{p}(\epsilon)} \vert\boldsymbol{\Phi}\vert/Q,&\text{for} \ \text{max-sum}  \\ 
		\min\limits_{k\in\mathbb{K}} R_k\cdot\min \limits_{\mathbf{p}(\epsilon)} \vert\boldsymbol{\Phi}\vert,&\text{for} \ \text{max-min}.
	\end{cases}
\end{equation}
where $\mathbf{p}(\epsilon)$ denotes the parametric expression of the moving target trajectory. $Q = card(\epsilon)$ is the sample point number of trajectory.

%\begin{equation}
%	\begin{array}{ll}
%		\max\limits_{\mathbf{t}, \mathbf{r}} & \left(\min\limits_{k\in\mathbb{K}} R_k, \min \limits_{\mathbf{p}(t)} \vert\boldsymbol{\Phi}\vert \right) \\
%		\text { s.t. } &x_{\text{min}} \leq x_i(x) \leq x_{\text{min}}, \quad i=1, \ldots, M+N \\
%		& y_{\text{min}} \leq y_i(x) \leq y_{\text{min}}, \quad i=1, \ldots, M+N,
%	\end{array}
%\end{equation}
%\begin{equation}
%	\begin{array}{ll}
%		\max\limits_{\mathbf{t}, \mathbf{r}} & 
%		\begin{cases}
%			\sum_{k=1}^{K}R_k\cdot\sum \limits_{\mathbf{p}(t)} \vert\boldsymbol{\Phi}\vert \\ 
%			\min\limits_{k\in\mathbb{K}} R_k\cdot\min \limits_{\mathbf{p}(t)} \vert\boldsymbol{\Phi}\vert
%		\end{cases}
%		\\
%		\text { s.t. } &x_{\text{min}} \leq x_i(x) \leq x_{\text{min}}, \quad i=1, \ldots, M+N \\
%		& y_{\text{min}} \leq y_i(x) \leq y_{\text{min}}, \quad i=1, \ldots, M+N,
%	\end{array}
%\end{equation}

\section{Soft Actor Critic Based AP Deployment}
Considering the original problem involving the multiplication of multiple objective functions is a complex non-convex optimization problem, traditional mathematical methods struggle to achieve the optimal solution. Thus, we first transform the problem into a Markov decision process (MDP) and utilize SAC for its solution. Compared to traditional deep reinforcement learning algorithms, SAC introduces a maximum entropy term in the loss function, which enhances its exploratory performance and mitigates the risk of falling into local optimal solution. This characteristic makes SAC particularly well-suited for addressing the complex multi-objective AP deployment problem.
\subsection{MDP Formulation}
Our MDP is represented by a tuple $\left( {{\mathbb S},{\mathbb A},P,R} \right)$, compared with traditional SAC, both state $\mathbb{S}$ and action $\mathbb{A}$ are discretized here to reduce the complexity of the exploration space and accelerate training procedure. $P:{\mathbb S} \times {\mathbb S} \times {\mathbb A} \to \left[ {0,\infty } \right)$ is the state transition probability density function to the next state ${\mathbf{s}_{i+1}} \in \mathbb{S}$ given current state ${\mathbf{s}_{i}} \in \mathbb{S}$ and action ${\mathbf{a}_{i}} \in \mathbb{A}$. After taking action, the agent gets reward $r(\mathbf{s}_i,\mathbf{a}_i) \in \mathbb{R}$ according to the reward function $R:{\mathbb S} \times {\mathbb A} \to \mathbb{R}$. The specific designs of states, actions, and rewards in our APs deployment problem are as follows.
\begin{itemize}
	\item State: Since the optimal deployment positions of APs are greatly relying on user and target positions, the state space is defined by the two-dimensional coordinates of users and targets.
	\begin{equation}
		\mathbf{s}_i = \left\{ {{\bf{u}}_1}, \ldots ,{{\bf{u}}_K},{\bf{p}} \right\}.
	\end{equation}
	\item Action: As mentioned in Section II, in our cell-free ISAC networks, the communication capacity and sensing accuracy are both related to the locations APs. Intuitively, we consider representing the action space using the 2D coordinates of APs. 
	\begin{equation}
		\mathbf{a}_i = \left\{ {{\mathbf{t}}_1^{i}}, \ldots ,{{\mathbf{t}}_M^{i}},{{\mathbf{r}}_1^{i}}, \ldots ,{{\bf{r}}_N^{i}} \right\}.
	\end{equation}
	\item Reward: The form of the reward function is determined by the expression of the objective function.
	\begin{equation}
		r(\mathbf{s}_i, \mathbf{a}_i) =
	\begin{cases}
	\sum_{k=1}^{K}R_k\cdot\sum \limits_{\mathbf{p}(\epsilon)} \vert\boldsymbol{\Phi}\vert,&\text{for} \ \text{max-sum}  \\ 
	\min\limits_{k\in\mathbb{K}} R_k\cdot\min \limits_{\mathbf{p}(\epsilon)} \vert\boldsymbol{\Phi}\vert,&\text{for} \ \text{max-min}.
\end{cases}
	\end{equation}
\end{itemize}
\subsection{SAC for Deployment}
Traditional DRL is aiming to learn the optimal policy $\pi$ which maximizes the cumulative expected rewards, whereas SAC adds a maximum policy entropy term in the DRL objective function 
\begin{equation}\label{RL}
	\rm{C}(\pi)=\sum\nolimits_i {{\rm{E}_{(\mathbf{s}_i,\mathbf{a}_i) \sim {\rho _\pi }}}\left[ r\left( {\mathbf{s}_i,\mathbf{a}_i} \right) +\omega \mathcal{H}\left(\pi\left(\cdot \mid \mathbf{s}_i\right)\right) \right]},
\end{equation}
where $\pi$ represents the policy generated by actor network, $\rho_\pi$ is the historical state-action trajectory under the policy $\pi$, $\omega$ is a weighting parameter called temperature coefficient that balances the exploration and exploiting historical experience during the training procedure. ${\mathcal{H}}(\pi(\cdot|\mathbf{s}_i))=-\log \pi(\cdot|\mathbf{s}_i)$ is the entropy of action policy at state $\mathbf{s}_i$.
\begin{figure}[h]
	\centering
	\includegraphics[width=\linewidth]{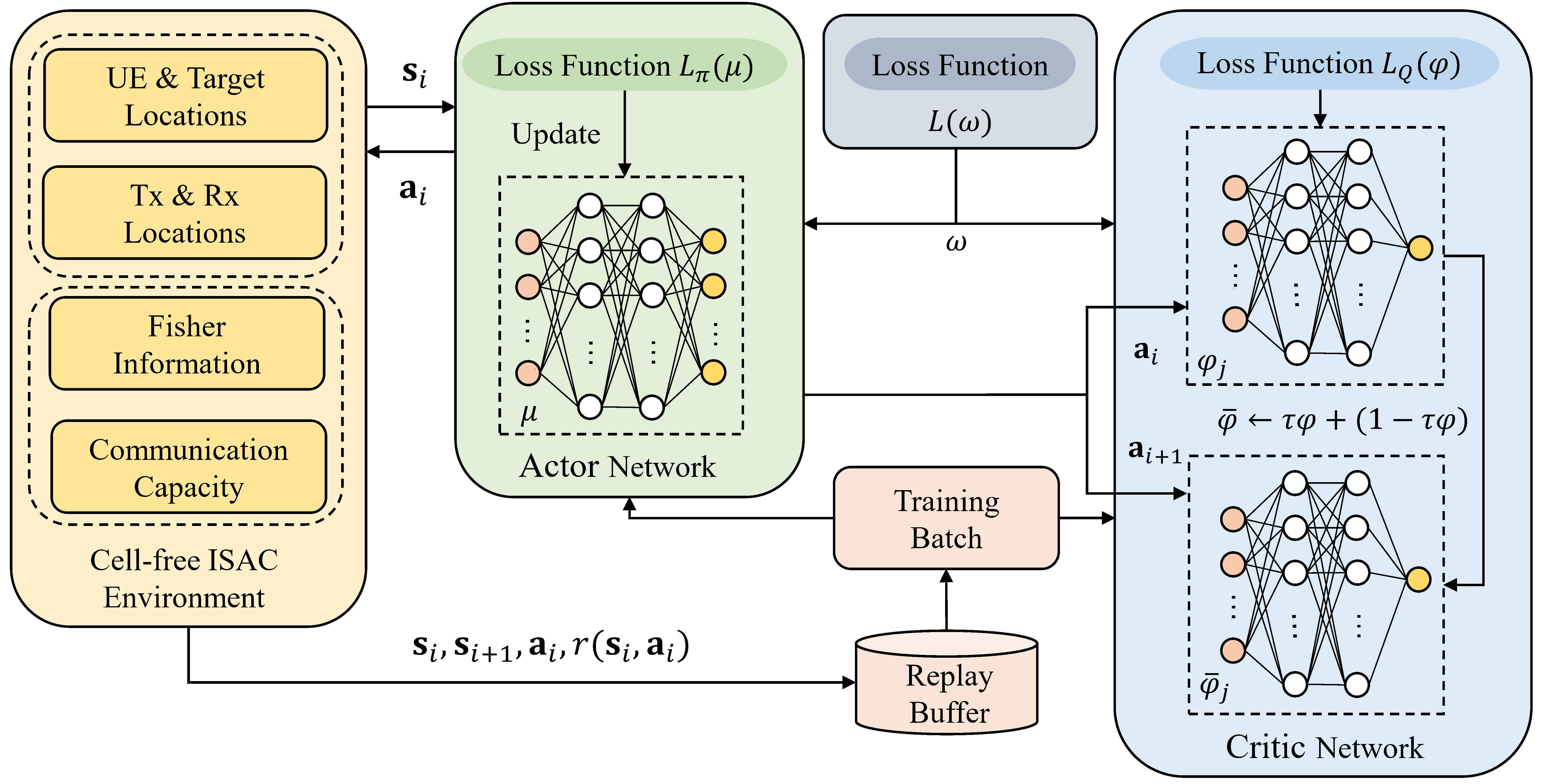}
	\caption{Architecture of SAC-based cell-free ISAC APs deployment system.}
	\label{SAC}
\end{figure}

As shown in Fig.~\ref{SAC}, the architecture of SAC-based cell-free ISAC APs deployment systems mainly contains three component. The first component is the training environment, responsible for receiving and executing actions from the actor network. Upon completion of execution, it generates the reward value for current state-action pair and transits to the next state, while storing corresponding state transition historical experience in the replay buffer $\mathcal{B}$ for neural network updates. The second part is the SAC actor network, denoted by parameters $\mu$, with its loss function defined as follows
\begin{equation}
	L_{\pi}(\mu)=\rm{E}_{\mathbf{s}_i \sim \mathcal{B}}\left[\rm{E}_{\mathbf{a}_i \sim \pi_\mu}[\omega \log(\pi_\mu(\mathbf{a}_i|\mathbf{s}_i)) - Q_\varphi({\mathbf{s}}_i,{\mathbf{a}}_i)]\right].
\end{equation}

The actor network generates action vectors corresponding to input state vectors and feeds them back to the environment for execution. Additionally, the critic network also requires the action vector when updating the Bellman equation.

The third component is the critic network parameterized by $\varphi$, with the loss function expressed as follows
\begin{equation}
	\begin{aligned}
		L_Q(\varphi) & =  \rm{E}_{\left(\mathbf{s}_i, \mathbf{a}_i\right)\sim\mathcal{B}} \left[ \frac{1}{2}\left(Q_\varphi(\mathbf{s}_i,\mathbf{a}_i)-\left(r(\mathbf{s}_i,\mathbf{a}_i)\right.\right.\right.\\
		& \left.\left.\left.+ \gamma Q_{\bar \varphi}\left(\mathbf{s}_{i+1}, \mathbf{a}_{i+1}\right)-\log \pi_\mu\left(\mathbf{a}_{i+1} \mid \mathbf{s}_{i+1}\right)\right)\right)^2 \right],
	\end{aligned}
\end{equation}
where $\gamma$ is discount factor. 

The critic network takes a state-action pair as input and outputs a Q-value used to assess the quality of the action taken in the current state, where a higher Q-value indicates the potential for greater cumulative reward. In contrast to the original critic network $\varphi$, the target critic network $\bar \varphi$ generates state-action pair for the next state, which is used in updating the Bellman equation. To mitigate training instability caused by overly large Q-value estimates, both $\varphi_j$ and $\bar \varphi_j$, $j\in\{1,2\}$ maintain two networks and utilize the smaller Q-value for network updates. A soft update strategy $\bar \varphi \leftarrow \tau \varphi + (1-\tau) \bar \varphi $ with $\tau \ll 1$ is employed.

In addition to these three main components, another important parameter, the temperature coefficient $\omega$, is also automatically updated according to its loss function
\begin{equation}
	L(\omega) = \rm{E}_{\mathbf{a}_i\sim\pi_{\mu}}\left[-\omega_i\log\pi_{\mu}(\mathbf{a}_i|\mathbf{s}_i)-\omega_i\overline{\mathcal{H}}\right],
\end{equation}
where $\overline{\mathcal{H}} = -|\mathcal{A}|_{\rm{dim}}$ represents the predefined lower bound of the action policy entropy.

\section{Simulation}
\begin{table}[h]
	\caption{Parameters in DRL model}
	\label{tab:DRL}
	\setlength{\tabcolsep}{20pt}
	\begin{tabular}{cc}
		\toprule
		Parameters & values \\
		\midrule
		Hidden layer & 64$\times$32\\
		Learning rate & $10^{-5}$\\
		Buffer size & $2^{21}$\\
		Batch size & $2^{9}$\\
		Discount factor $\gamma$ & 0.98\\
		Soft update target critic network $\tau$ & 0.005\\
		\bottomrule
	\end{tabular}
\end{table}
%\begin{table}[h]
%	\caption{Parameters in DRL model}
%	\label{tab:DRL}
%	\begin{tabular}{ccccccc}
%		\toprule
%		Parameters & Hidden layer & Learning rate & Buffer size & Batch size & $\gamma$ & $\tau$ \\
%		\midrule
%		Values & 64$\times$32 & $10^{-5}$ & $2^{21}$ & $2^{9}$ & 0.98 & 0.005\\
%		\bottomrule
%	\end{tabular}
%\end{table}
In this section, we evaluated the performance of our proposed SAC-based APs deployment method in cell-free ISAC systems. We demonstrated the superiority of SAC in this scenario through comparisons with several traditional DRL algorithms. Additionally, to validate the effectiveness of optimizing the joint communication and sensing metric proposed in this work, we compared it with results of  only sensing optimizing, only communication optimizing, and weighted sum of communication and sensing optimizing. The configuration of DRL Model is show in Table~\ref{tab:DRL}. In our cell-free systems, the target moves along a predefined circular trajectory, while three UEs distributed around the target follow Gaussian distributions with a variance of 2. During sensing tasks, receiving APs capture echoes transmitted from transmitting AP, reflected from the target. During communication tasks, all APs jointly receive uplink signals from all UEs.
\begin{figure}[h]
	\centering	
		\subfigure[]{
		\centering
		\includegraphics[scale=0.45]{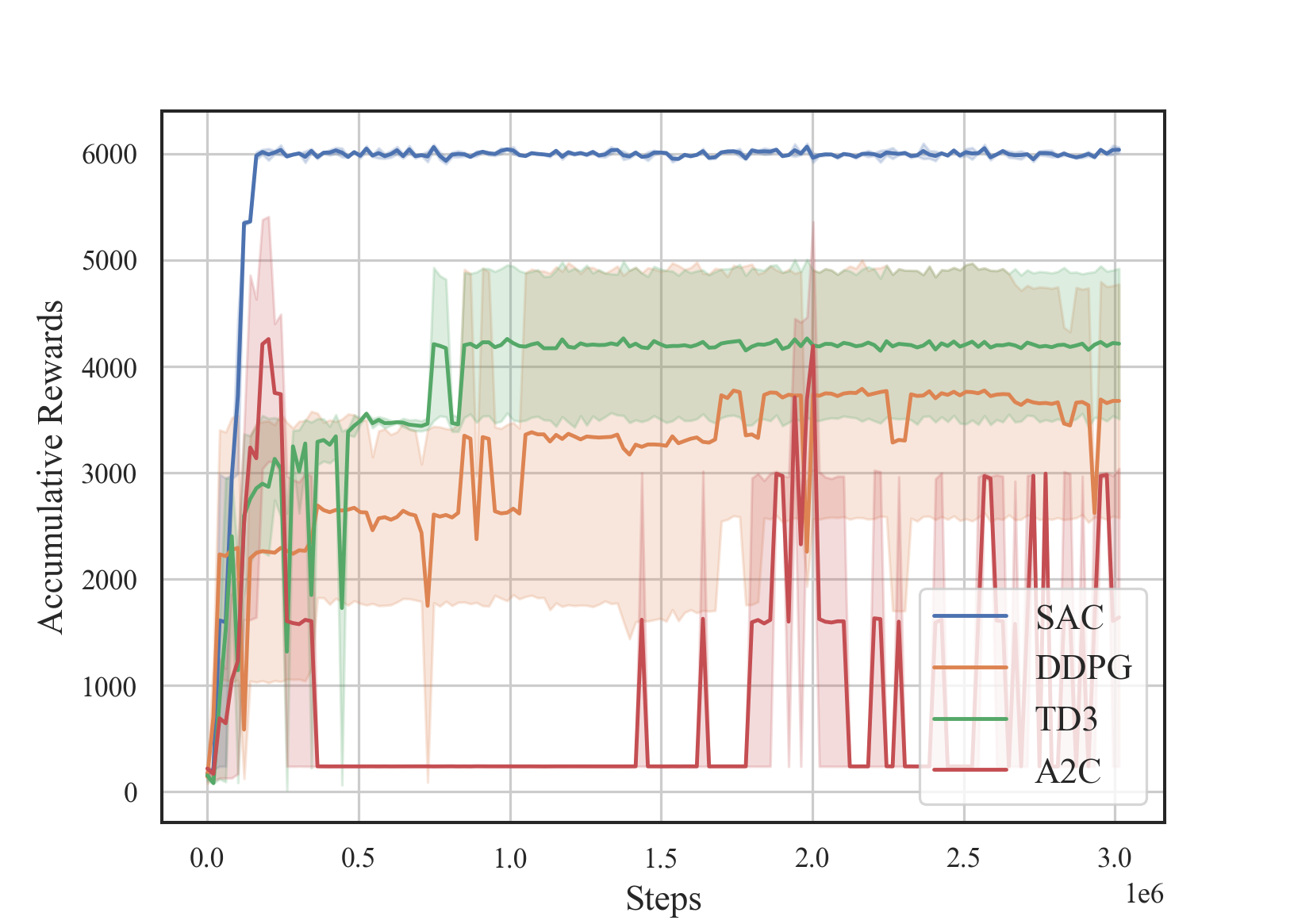}
		\label{algo.sub.1}
	}\vspace{-4mm}
	\subfigure[]{
		\centering
		\includegraphics[scale=0.45]{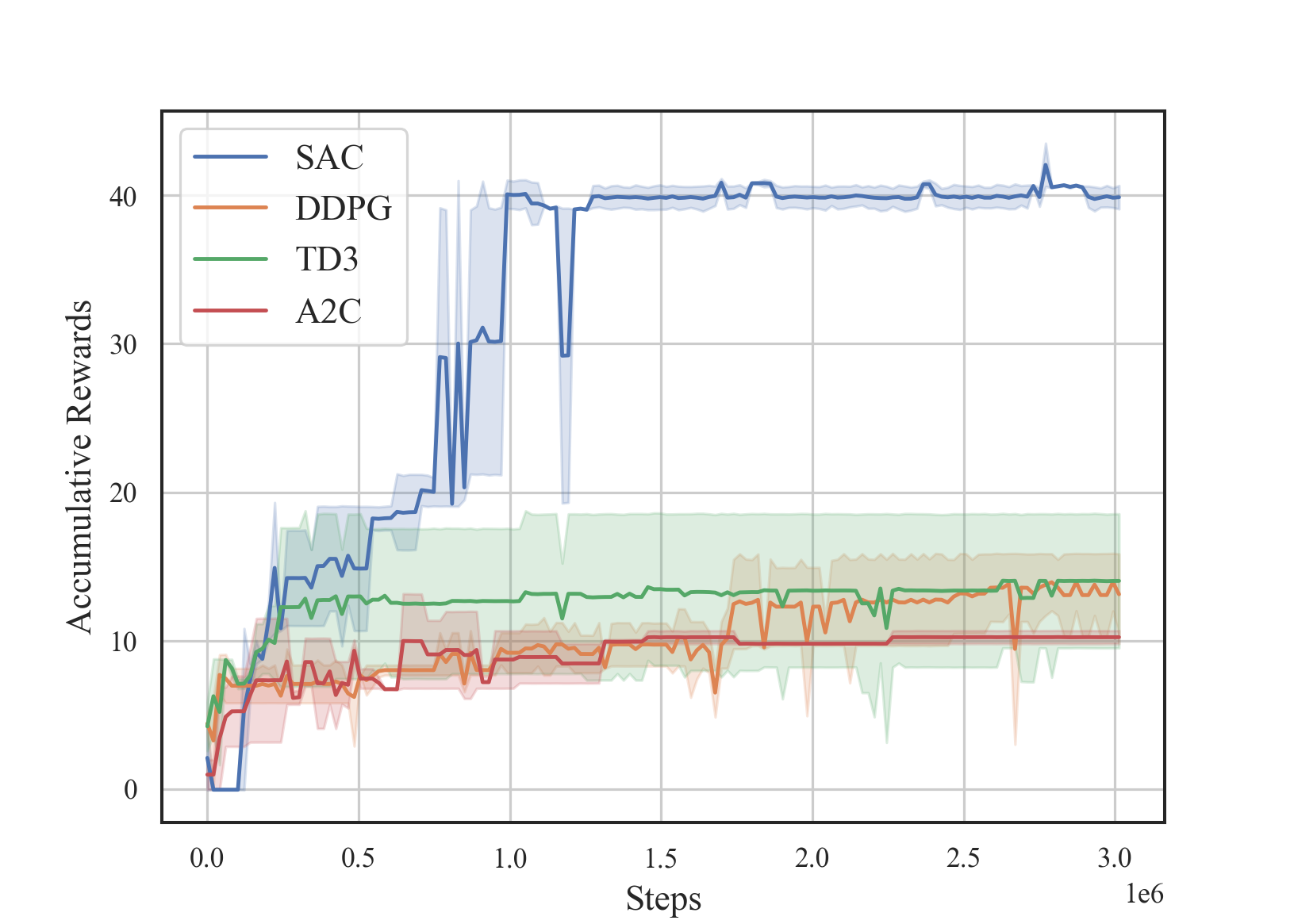}
		\label{algo.sub.2}
	}
	\caption{Accumulative ISAC rewards of different DRL Algorithms. (a) Max-sum problem. (b) Max-min problem.}
	\label{simu_algorithm}
\end{figure}

 As shown in Fig.~\ref{simu_algorithm}, in cell-free ISAC systems, the final convergence result of our proposed SAC-based algorithm significantly surpasses other traditional DRL algorithms in both max-min and max-sum problem. Also, SAC is more stable under different random seeds compared  with other algorithms. The hyperparameters of DDPG and TD3 exert a profound impact on their performance, typically, for a complex optimization problem, finding hyperparameters that enable them to exceeds the performance of SAC is challenging. According to the data in Table~\ref{apnumber}, an increase in the number of APs is observed to correlate positively with improved ISAC performance, thereby significantly demonstrating the pronounced advantage of the cell-free systems in this regard.
 
\begin{figure}[h]
	\centering	
	\subfigure[SAC]{
		\includegraphics[scale=0.43]{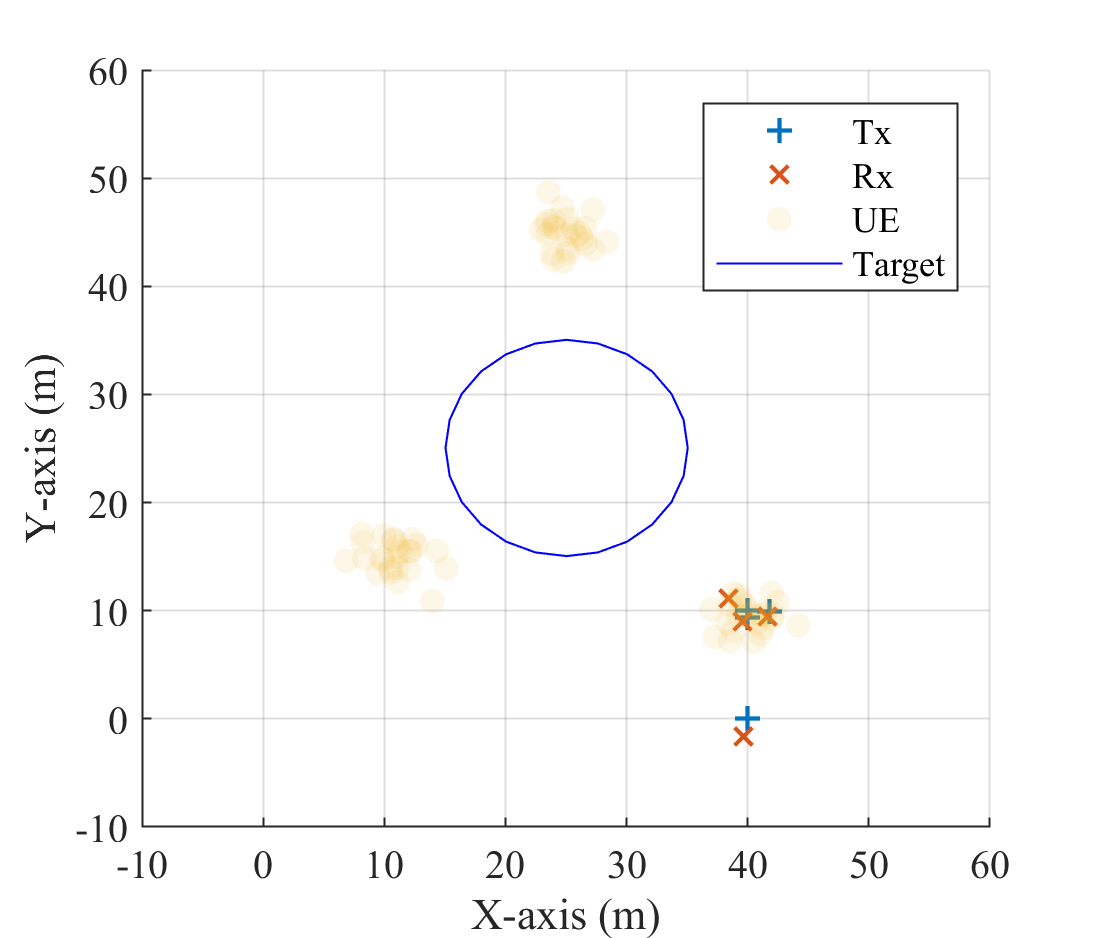}
		\label{dep.sub.1}
	}\vspace{-4mm}
	\subfigure[TD3]{
		\includegraphics[scale=0.43]{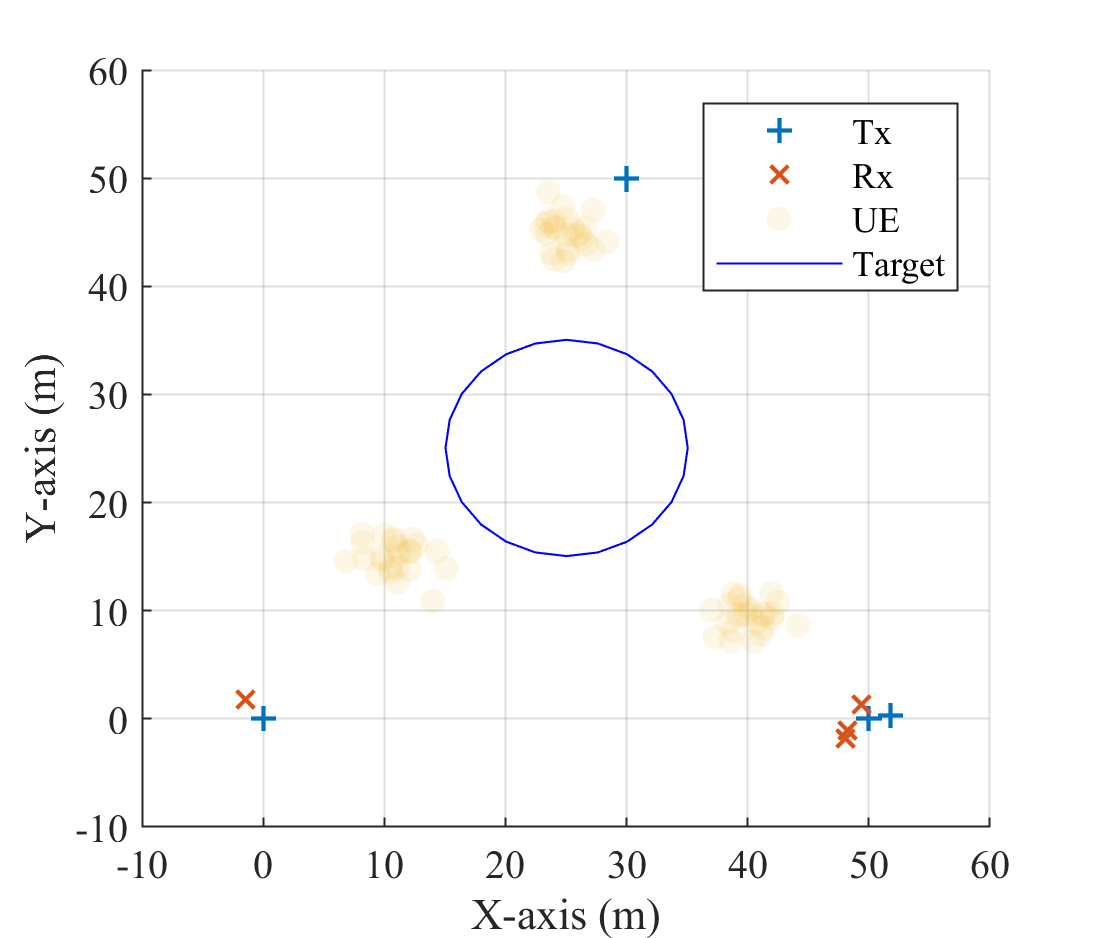}
		\label{dep.sub.2}
	}\\
	\subfigure[DDPG]{
	\includegraphics[scale=0.43]{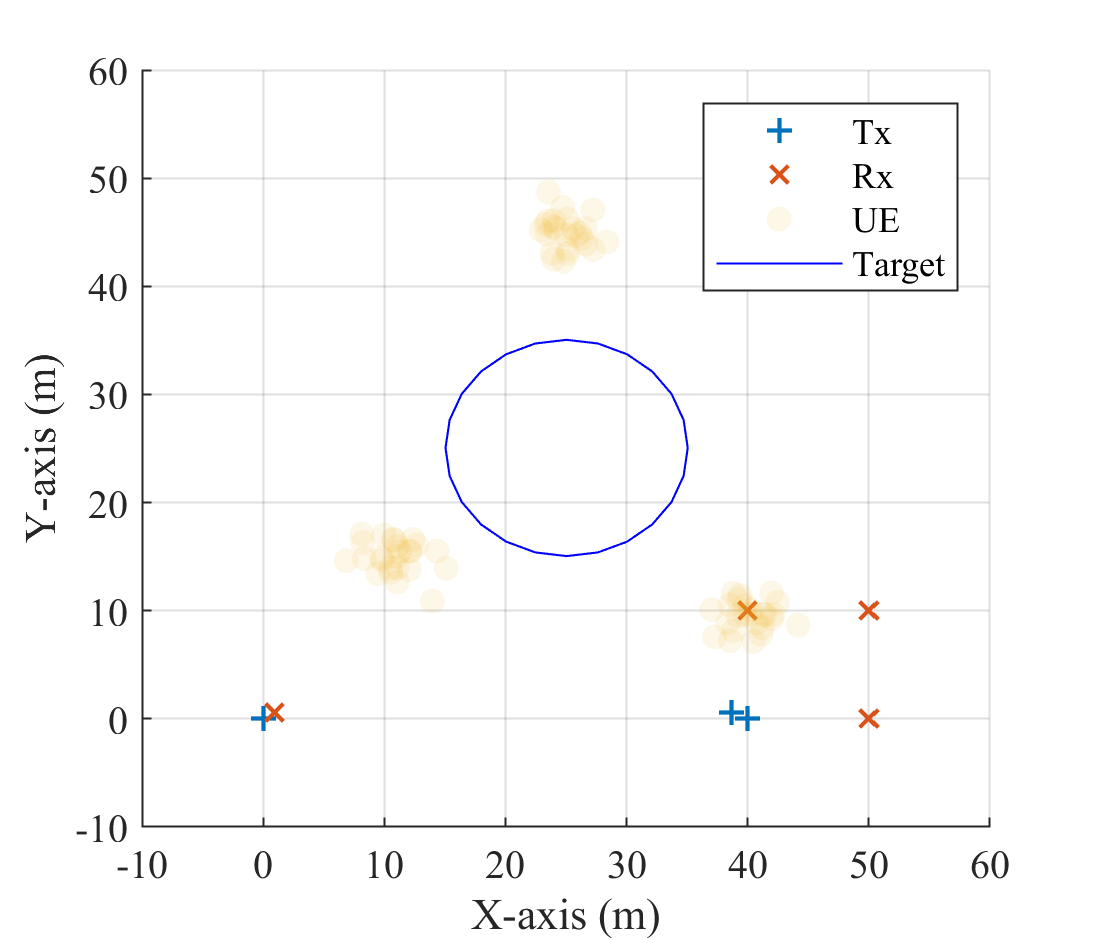}
	\label{dep.sub.3}
}
		\subfigure[A2C]{
		\includegraphics[scale=0.43]{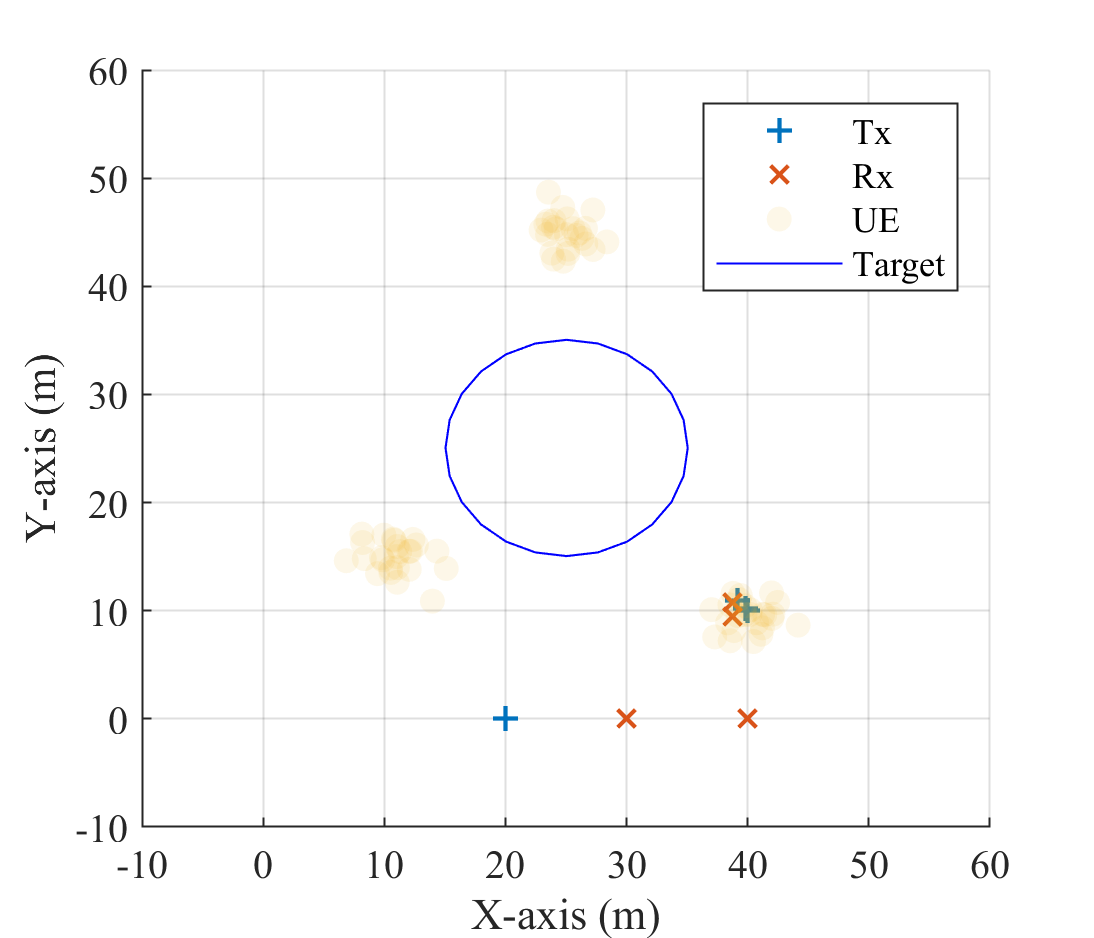}
		\label{dep.sub.4}
	}\\
		\subfigure[Communication only]{
		\includegraphics[scale=0.43]{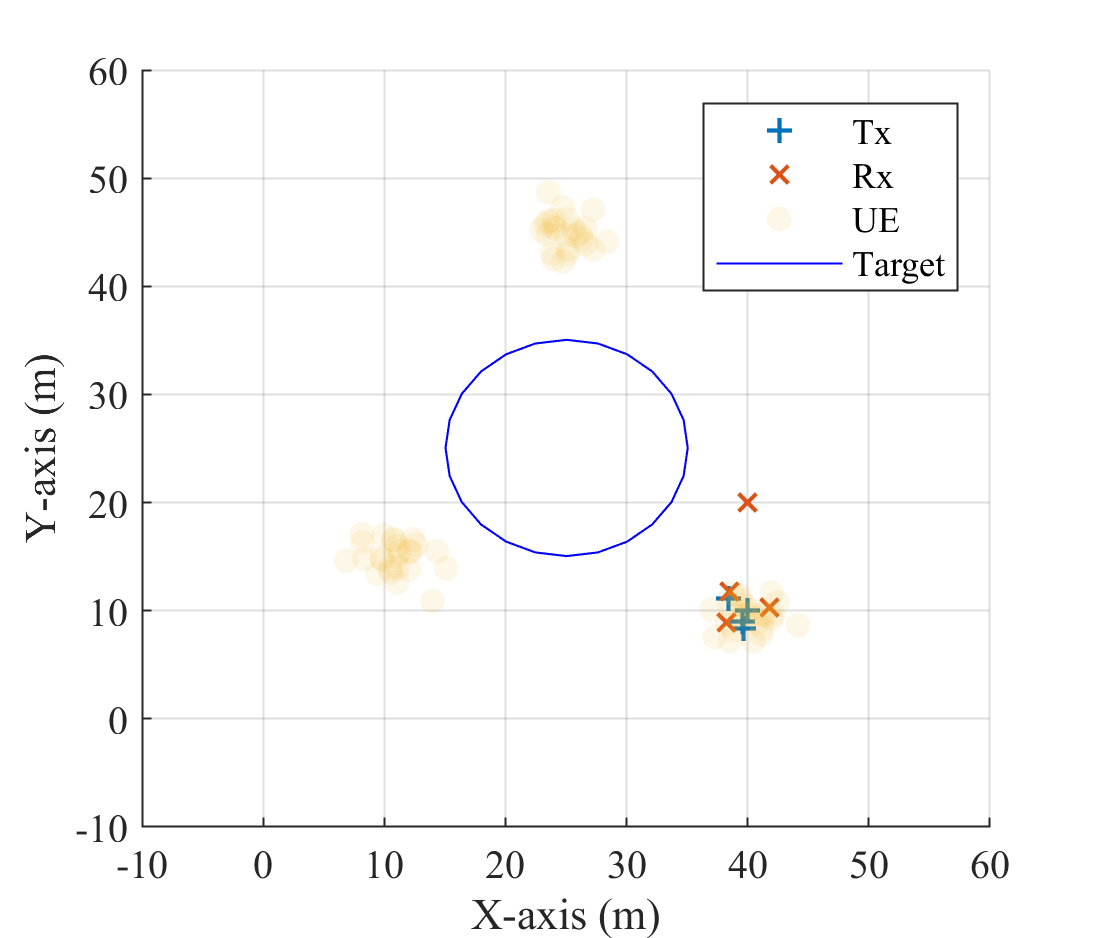}
		\label{dep.sub.5}
	}
	\subfigure[Sensing only]{
		\includegraphics[scale=0.43]{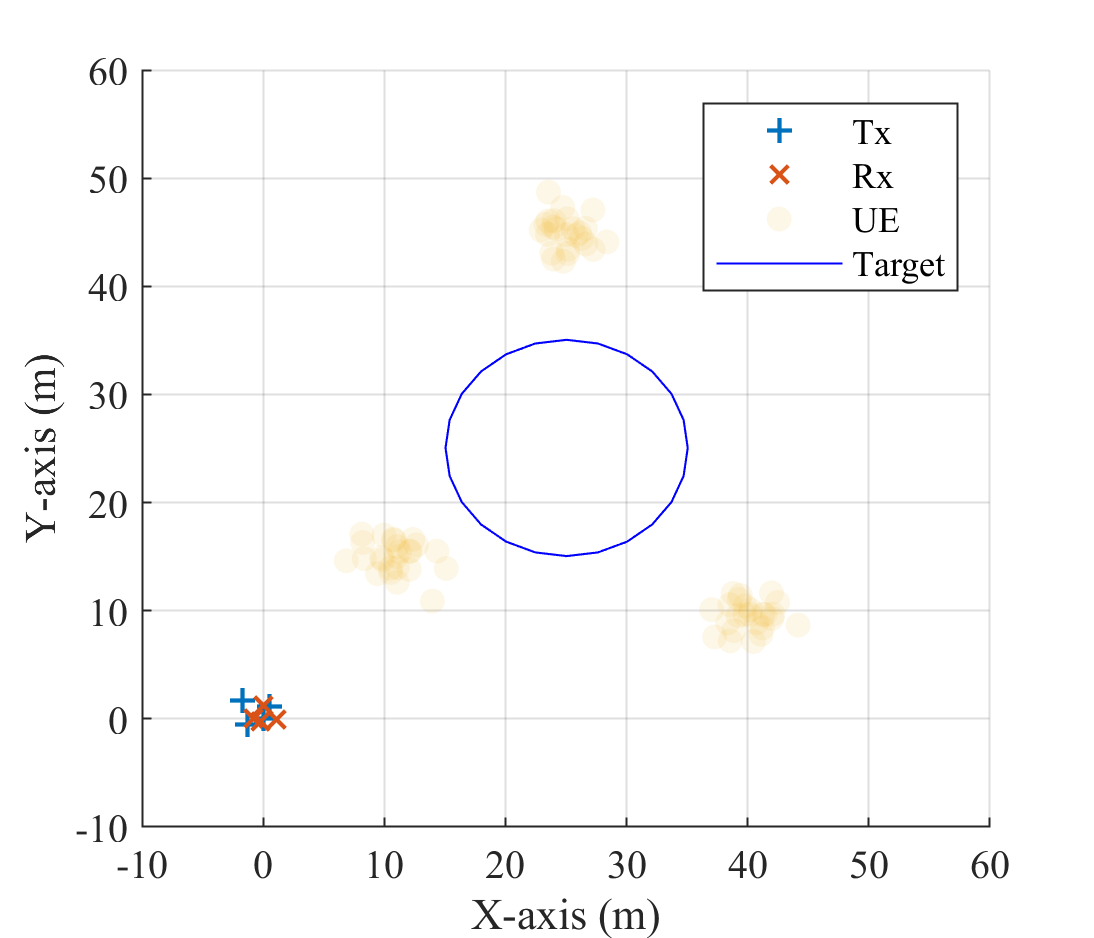}
		\label{dep.sub.6}
	}	
	\caption{APs deployment results of max-sum problem.}
	\label{dep_maxmean}
\end{figure}
\begin{figure}[h]
	\centering
			\subfigure[SAC]{
		\includegraphics[scale=0.43]{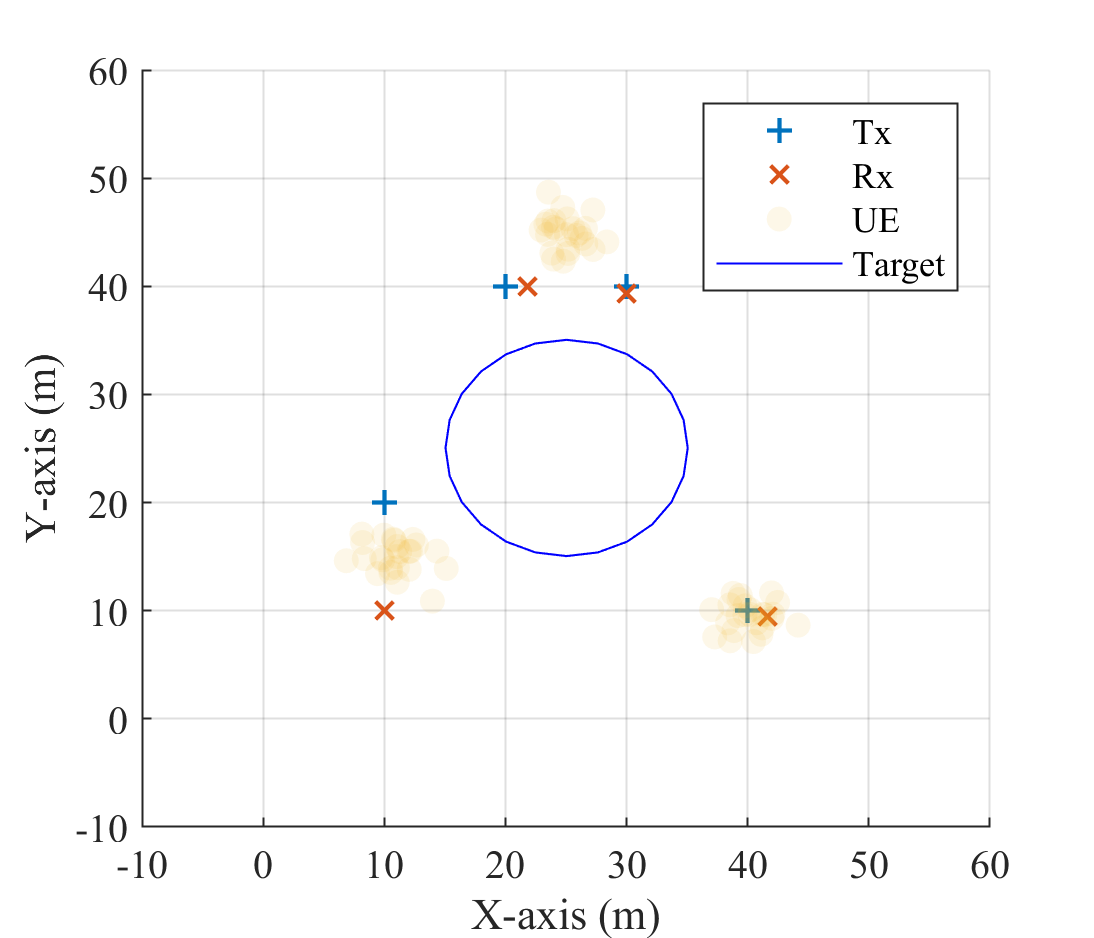}
		\label{dep.sub.7}
	}
	\subfigure[TD3]{
		\centering
		\includegraphics[scale=0.43]{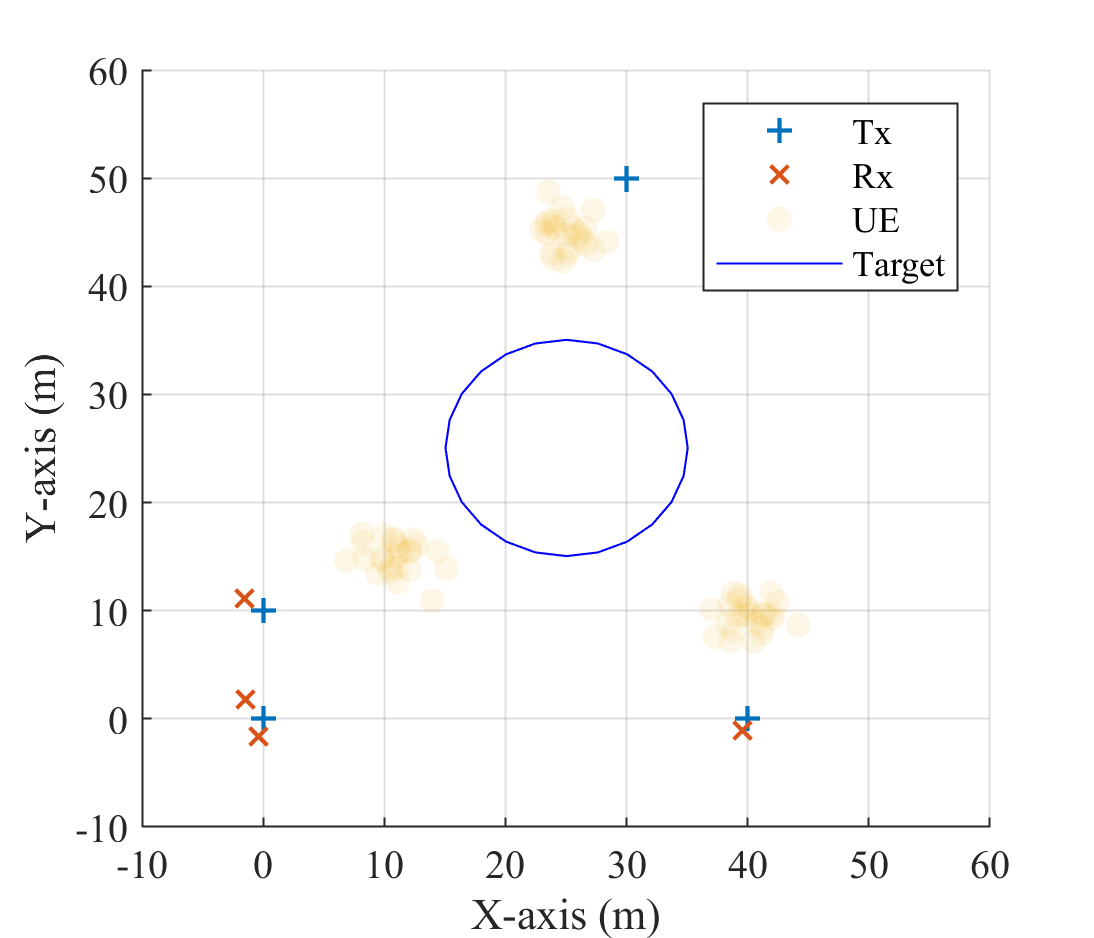}
		\label{dep.sub.8}
	}\\
	\subfigure[DDPG]{
		\includegraphics[scale=0.43]{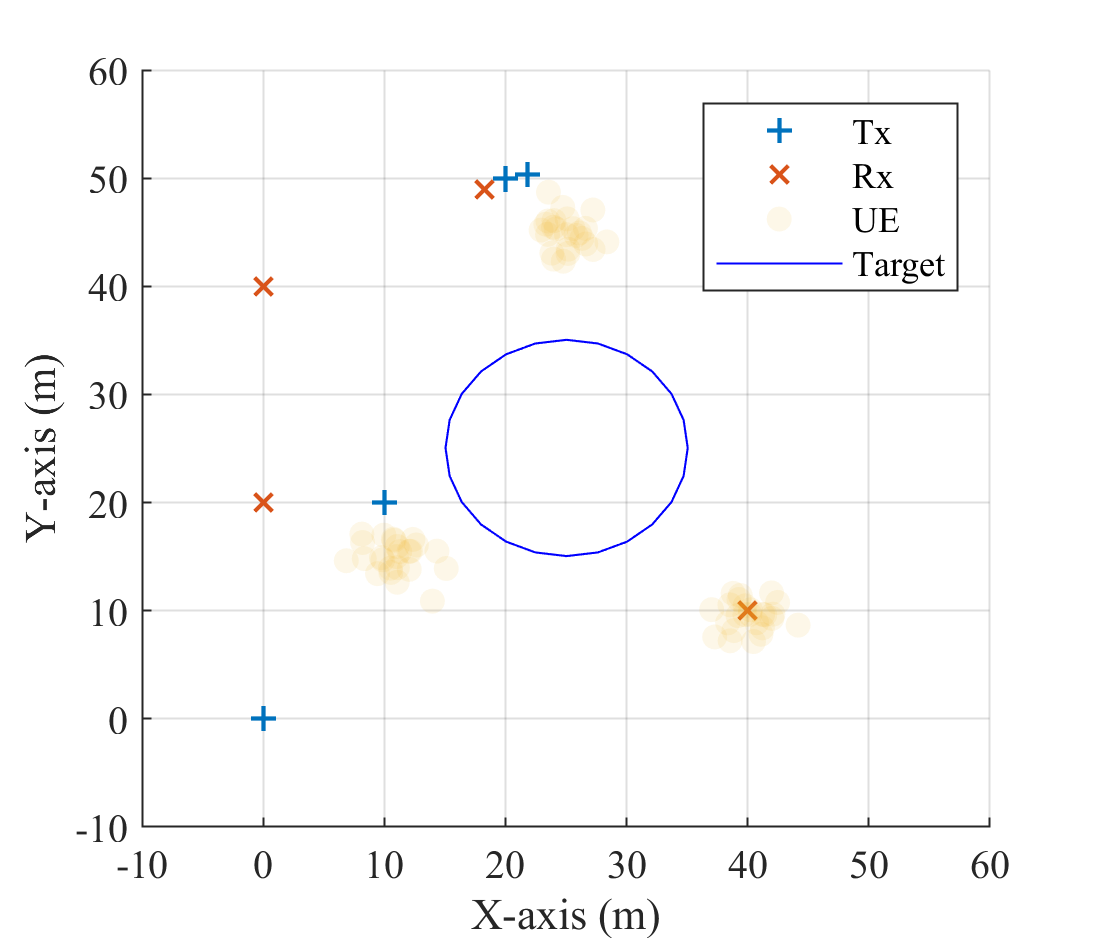}
		\label{dep.sub.9}
	}
	\subfigure[A2C]{
		\includegraphics[scale=0.43]{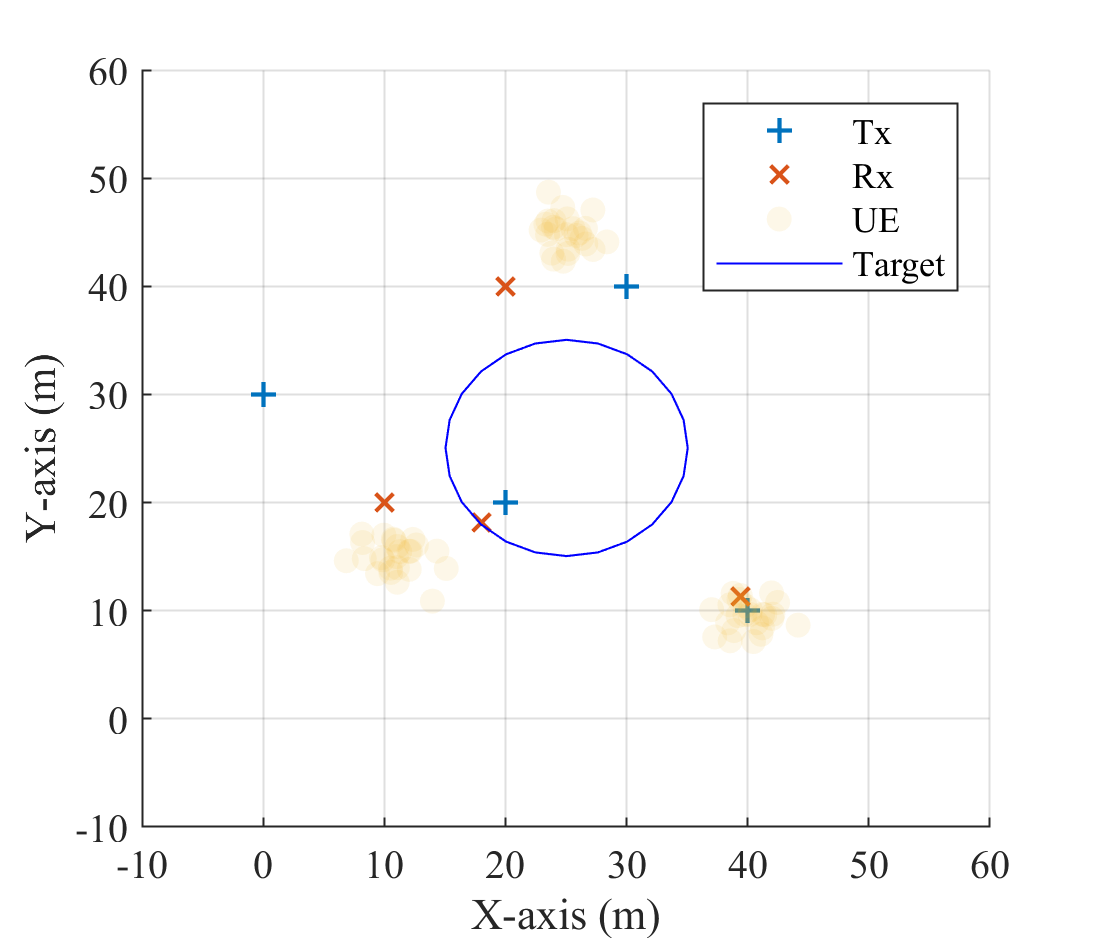}
		\label{dep.sub.10}
	}\\
	\subfigure[Communication only]{
		\includegraphics[scale=0.43]{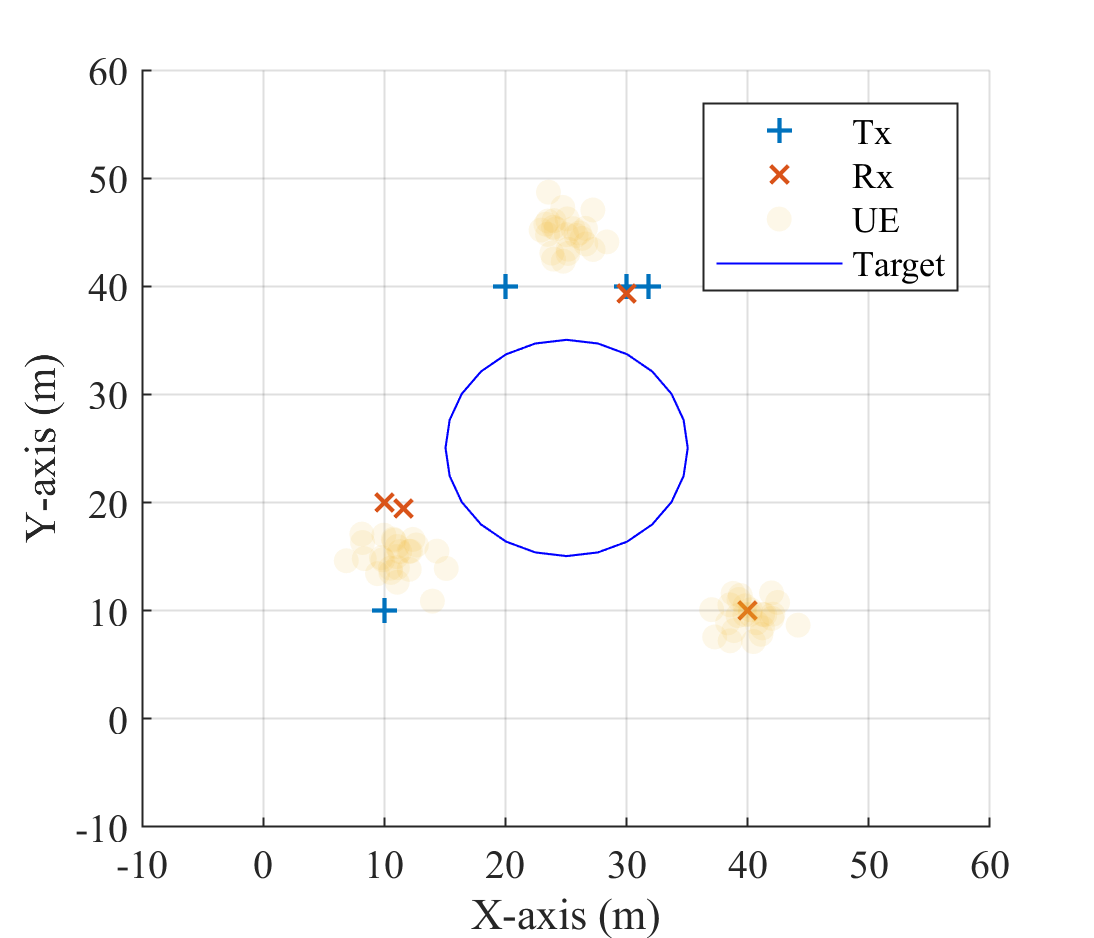}
		\label{dep.sub.11}
	}
	\subfigure[Sensing only]{
		\includegraphics[scale=0.43]{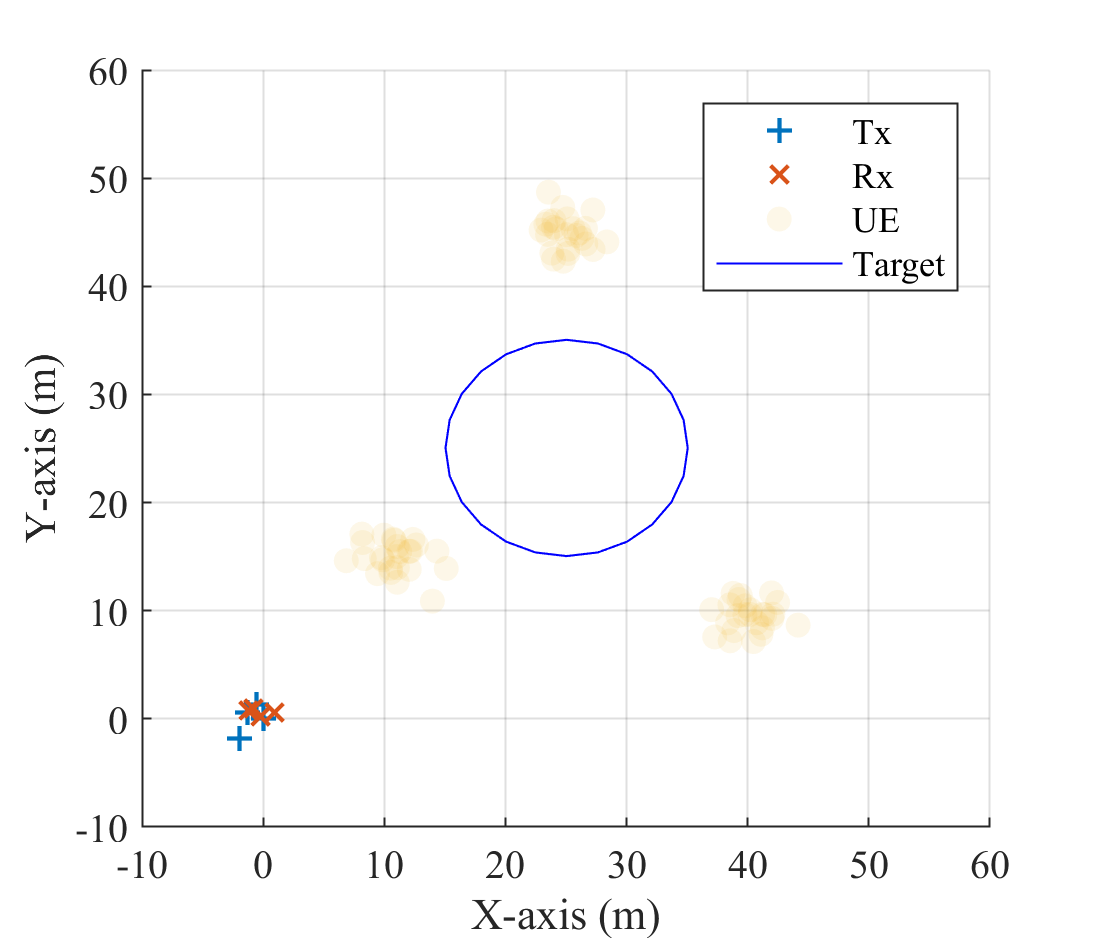}
		\label{dep.sub.12}
	}
	\caption{APs deployment results of max-min problem.}
	\label{dep_maxmin}
\end{figure}
%\begin{figure}[h]
%	\centering
%	\includegraphics[width=\linewidth]{maxmean_isac_dep}
%	\caption{Max-sum APs deployment result in cell-free ISAC system}
%	\label{max-sum dep}
%\end{figure}
%
%\begin{figure}[h]
%	\centering
%	\includegraphics[width=\linewidth]{maxmin_isac_dep}
%	\caption{Max-min deployment result in cell-free ISAC system}
%	\label{max-min dep}
%\end{figure}
\begin{table}[h]
	\centering
	\caption{ISAC value of different number of APs}
	\label{apnumber}
	\setlength{\tabcolsep}{3pt}
	\begin{tabular}{lcccc}
		\toprule
		Number of APs (M+N) & 2 & 4 & 6 & 8 \\
		\midrule		
		$\sum_{k=1}^{K}R_k/Q\cdot\sum \limits_{\mathbf{p}(\epsilon)} \vert\boldsymbol{\Phi}\vert/Q$ & 41.07 & 775.73 & 3804.47 & 11614.07 \vspace{2mm} \\
		$\min\limits_{k\in\mathbb{K}} R_k\cdot\min \limits_{\mathbf{p}(\epsilon)} \vert\boldsymbol{\Phi}\vert$ & 0.042 & 10.48 & 86.06 & 257.28\\
		\bottomrule
	\end{tabular}
\end{table}
\begin{table}[h]
	\centering
	\caption{Communication rate and localizing accuracy of different objective}
	\label{objective}
	\setlength{\tabcolsep}{3pt}
		\begin{tabular}{lcccc}
			\toprule
			Objective & ISAC & Sensing only & Comm. only & Weighted sum \\
			\midrule

			$\sum_{k=1}^{K}R_k/Q$ & 1.13 & 0.014 & 1.26 & 0.015 \vspace{2mm} \\
			$\sum \limits_{\mathbf{p}(\epsilon)} \vert\boldsymbol{\Phi}\vert /Q
			$ & 688.97	& 842.92 & 642.08 & 842.92 \\
			$\min\limits_{k\in\mathbb{K}} R_k$ & 0.042 & 0.0013 & 0.043 & 0.0013\\
			$\min \limits_{\mathbf{p}(\epsilon)} \vert\boldsymbol{\Phi}\vert$ & 0.25 & 696.89 & 188.40 & 696.89\\
			\bottomrule
		\end{tabular}
\end{table}

To be more specific, Fig.~\ref{dep_maxmean} and Fig.~\ref{dep_maxmin} depict the APs deployment results of the cell-free ISAC systems using different algorithms in max-sum and max-min problems. According to the D-optimal criterion, optimal sensing performance is achieved when the positions of transmitting APs and receiving APs lie on a straight line and form specific angles with the target. According to the Euclidean distance criterion derived from zero-forcing reception, communication performance improves as APs approach to UEs. SAC-based APs deployment result satisfies both sensing and communication criterion well. Furthermore, APs deployment in max-min problem in Fig.~\ref{dep.sub.7} shows that to ensure fairness in service among UEs, each UE is surrounded by several nearby APs to achieve higher SNR. In contrast, in the deployment result of the max-sum problem in Fig.~\ref{dep.sub.1}, all APs are concentrated around the UEs closest to the optimal sensing position. 
%\begin{figure}[h]
%	\centering
%	\includegraphics[width=\linewidth]{metric_bar}
%	\caption{Communication and sensing values of different objective functions}
%	\label{metric_bar}
%\end{figure}

Table~\ref{objective} illustrates the communication and sensing values under optimal APs deployment in different objective functions. In both max-min and max-sum problem, the communication values of our ISAC optimization is almost equal to the result of only optimizing the communication objective, and far higher than only optimizing sensing method, while the sensing values of ISAC optimization is better than communication method but lower than sensing-only method. This is because the optimal APs deployment position for sensing are typically aligned in a straight line at specific angles with the target. However, UEs distribution usually does not conform to this characteristic. Therefore, to balance communication rate of UEs, APs cannot be deployed at the optimal sensing position, resulting in a sacrifice of some sensing performance. In addition, due to the significantly different magnitudes between sensing accuracy and communication rate, finding a balanced weight in their sum as a objective function proves challenging. It can be observed that the result of weighted sum method is nearly equivalent to optimizing sensing performance alone. 
%\begin{figure}[h]
%	\centering
%	\includegraphics[width=\linewidth]{metric_dep}
%	\caption{APs deployment result of different objective functions}
%	\label{metric_dep}
%\end{figure}

Fig.~\ref{dep.sub.6} and Fig.~\ref{dep.sub.12} presents the deployment results of optimizing communication alone and optimizing sensing alone. As previously analyzed, for optimal sensing performance, APs form a line with target at a specific angles. For optimal communication performance, as shown in Fig.~\ref{dep.sub.5} and Fig.~\ref{dep.sub.11}, APs strive to be as close to the UEs as possible to achieve higher SNR.
\section{Conclusion}
In this work, we have firstly investigated the APs deployment problem for maximizing user rate and localizing accuracy in the cell-free ISAC systems. Then a unified evaluation metric merging D-optimal criterion with Euclidean distance have been designed, which enables simultaneous optimization of communication and sensing performance. Finally, we have employed SAC, a DRL algorithm augmented with maximum action policy entropy to solve the original non-convex and high-dimensional problem. Numerical findings demonstrated superior convergence results of the proposed method compared with other DRL methods for both system overall performance and fairness performance.
	
\begin{acks}
	This work was supported in part by the Fundamental Research Funds for the Central Universities under Grant Nos. 2242022k60002 and 2242023R40005.
\end{acks}


%\begin{acks}
%To Robert, for the bagels and explaining CMYK and color spaces.
%\end{acks}

\bibliographystyle{ACM-Reference-Format}
\bibliography{ref}

\appendix

\end{document}